\documentstyle[11pt,aaspp4]{article}

\slugcomment{Submitted to {\it ApJ Letters}, October 1996}

\begin{document}

\title{Effect of Gravitational Lensing on Measurements \\ of the
Sunyaev-Zel'dovich Effect} \author{Abraham Loeb$^{1,3}$ and Alexandre
Refregier$^{2,4}$}
\medskip
\affil{1. Astronomy Department, Harvard University, 60 Garden St.,
Cambridge, MA 02138}
\affil{2. Columbia Astrophysics Laboratory, 538 W. 120th Street,
New York, NY 10027}
\altaffiltext{3}{email:aloeb@cfa.harvard.edu}
\altaffiltext{4}{email:refreg@odyssey.phys.columbia.edu; also at the
Department of Physics, Columbia University}

\begin{abstract}
The Sunyaev--Zel'dovich (SZ) effect of a cluster of galaxies is usually
measured after background radio sources are removed from the cluster field.
Gravitational lensing by the cluster potential leads to a systematic
deficit in the residual intensity of unresolved sources behind the cluster
core relative to a control field far from the cluster center. As a result,
the measured decrement in the Rayleigh--Jeans temperature of the cosmic
microwave background is overestimated. We calculate the associated
systematic bias which is inevitably introduced into measurements of the
Hubble constant using the SZ effect.  For the cluster A2218, we find that
observations at 15 GHz with a beam radius of $0^{\prime}.4$ and a source
removal threshold of $100\mu{\rm Jy}$ underestimate the Hubble constant by
6--10\%.  If the profile of the gas pressure declines more steeply with
radius than that of the dark matter density, then the ratio of lensing to
SZ decrements increases towards the outer part of the cluster.
\end{abstract}

\keywords{cosmic microwave background -- diffuse radiation --
galaxies: clusters: general, individual (A2218) -- gravitational
lensing}

\section{Introduction}

The Sunyaev--Zel'dovich (SZ) effect describes the distortion introduced
to the Cosmic Microwave Background (CMB) spectrum due to its Compton
scattering off free electrons, which are either hot (the {\it thermal}
effect) or possess a bulk peculiar velocity (the {\it kinematic}
effect; see reviews of both effects in Sunyaev \& Zel'dovich 1980 and
Rephaeli 1995).  The thermal SZ effect provides an important
diagnostic of the hot gas in clusters of galaxies, and by now has been
measured in a number of systems (see Table 1 in Rephaeli 1995). The
kinematic effect has an amplitude which is typically an order-of
magnitude smaller and has not yet been definitively detected (see
Rephaeli \& Lahav 1991 and Haehnelt \& Tegmark 1995, regarding
prospects for a future detection).

It has long been realized that a measurement of the thermal SZ effect,
combined with X--ray observations, can be used to estimate the
distance to the cluster and hence the Hubble constant, $H_0$, under
the assumption that the cluster is spherical (Cavaliere, Danese, \& De
Zotti 1977, Gunn 1978, Silk \& White 1978, Birkinshaw 1979). The
inferred value of the Hubble constant is inversely proportional to the
square of the SZ temperature decrement.  This approach had led to
values of the Hubble constant which are typically on the low side of
the range inferred from other methods (see, e.g., Table 2 in Rephaeli
1995). An often cited systematic effect that could account for this
bias is elongation of the selected clusters along the line-of-sight.
In this {\it Letter}, we explore a different effect which leads to a
systematic bias towards low--$H_0$ values even if these clusters are
perfectly spherical.  The effect results from gravitational lensing by
the cluster potentials.

Measurements of the decrement in the Rayleigh--Jeans (RJ) temperature
of the microwave background due to the thermal SZ effect are routinely
accompanied by the removal of background radio sources down to some
flux threshold (see, e.g., Birkinshaw, Hughes, \& Arnaud 1991). In this
process, it is implicitly assumed that the flux threshold for the
removal of sources behind the cluster core is the same as in a control
field far from the cluster center.  However, this assumption is not
strictly true due to the inevitable magnification bias which is
introduced by the gravitational lensing effect of the cluster
potential.  In reality, the cluster acts as a lens which magnifies and
thus resolves sources that are otherwise below the detection
threshold.  The residual intensity of unresolved sources is therefore
systematically lower behind the cluster core, as compared to that in
the control field. Lensing artificially increases the flux deficit
behind the cluster core and thus leads to a systematic underestimate
of the Hubble constant.

In this {\it Letter} we calculate the effect of lensing on SZ measurements
of the Hubble constant. Our discussion on lensing follows closely the
approach developed in an earlier paper (\markcite{ref96}Refregier \& Loeb
1996, hereafter RL) which focused on lensing of the X--ray background by
galaxy clusters; the interested reader should consult this earlier paper
for more details. Here, we describe our models for the background
population of radio sources and for the cluster potential in \S\ref{model}.
We then show in
\S\ref{lensing_effect} how the lensing effect leads to a systematic
decrement in the intensity of unresolved sources.  In \S\ref{results},
we present numerical results for different values of our model
parameters and for the specific example of A2218. Finally,
\S\ref{conclusion} summarizes the main conclusions of this work.

\section{Model}
\label{model}

We model the gravitational potential of the cluster as a Singular
Isothermal Sphere (SIS) (e.g. \markcite{sch92}Schneider et al. 1992). This
model provides a good first--order approximation to the projected mass
distribution of known cluster lenses (Tyson \& Fischer 1995,
\markcite{nar96}Narayan \& Bartelmann 1996,
\markcite{sq96a}\markcite{sq96b}Squires et al. 1996a,b). The SIS
potential causes background sources to appear brighter but diluted on
the sky by the magnification factor 
\begin{equation}
\mu(\theta)=\left| 1 - \frac{\alpha}{\theta} \right|^{-1},
\label{eq:mu_theta}
\end{equation}
where $\theta$ is the angle between the image of the source and the lens
center, and $\alpha$ is the Einstein angle. For a SIS with a line-of-sight
velocity dispersion $\sigma_v$, $\alpha=4\pi(\sigma_v^2/c^2)D_{ls}/D_{os}$,
where $D_{ls}$ and $D_{os}$ are the lens--source and the observer--source
angular diameter distances, respectively.

In general, the Einstein angle $\alpha$ depends on the redshifts of the
lens, $z_{l}$, and of the source, $z_{s}$. However, the dependence on the
source redshift is weak if $z_s \ga 3z_{l}$ (cf. Fig. 1 in RL).  Most
measurements of the SZ effect are performed with nearby clusters ($z_{l}
\la 0.2$), while sub--mJy radio sources have median redshifts in the
range $0.5\la z_{s}\la 0.75$ (\markcite{win93}Windhorst et al.  1993).  
We therefore take $\alpha$ to be independent of $z_{s}$ and
simply consider the two--dimensional distribution of radio sources on the
sky.

\begin{figure}
\plotone{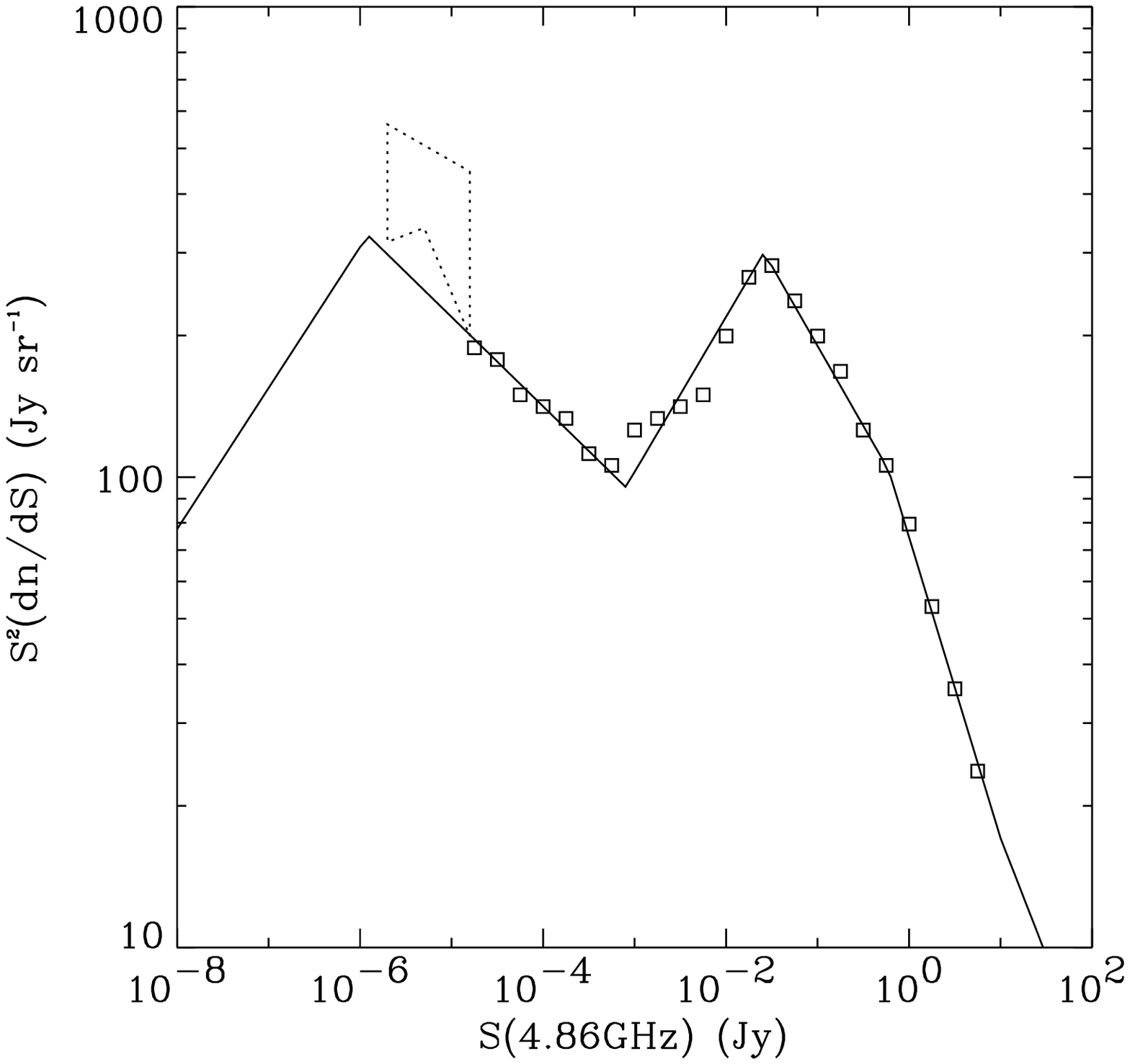}
\caption{Number--flux relation for radio sources at 4.86 GHz. The
counts were normalized to $S^{-2}$, the relation which remains
invariant under lensing.  The approximate mean counts summarized by
Windhorst et al.  (1993) are shown as squares. The dotted line
corresponds to the limits inferred from a fluctuation analysis of the
unresolved background (Fomalont et al. 1991). The solid line shows our
model with its six power--law components.
\label{fig:n_s_norm}}
\end{figure}

We model the flux distribution of background radio sources according to the
observed number--flux relation at 4.86 GHz (\markcite{win93}Windhorst et
al.  1993).  The squares in Figure~\ref{fig:n_s_norm} show the mean values
of the observed differential counts, ${dn}/{dS}$, normalized by $S^{-2}$.
The dotted line illustrates the number counts limits inferred by
\markcite{fom91}Fomalont et al. (1991) from a fluctuation analysis of
the radio source background (see also \markcite{win93}Windhorst et al.
1993). The observed counts extend from $\sim10^{-5}$ to 10 Jy.  We model
these counts by six broken power laws of the form, $\left.
({dn}/{dS})\right|_{S} = \eta_{j} S^{-\lambda_{j}}$ for
$S_{j-1,j}>S>S_{j,j+1}$, with $j=1, \dots ,6$. The six slope parameters are
$\lambda_{1}=2.50$, $\lambda_{2}=2.64$, $\lambda_{3}=2.34$,
$\lambda_{4}=1.67$, $\lambda_{5}=2.19$, and $\lambda_{6}=1.70$.  The
power-law break points are $S_{12}=10$, $S_{23}=6.0 \times 10^{-1}$,
$S_{34}=2.6 \times 10^{-2}$, $S_{45}=8.0
\times 10^{-4}$, and $S_{56}=1.2 \times 10^{-6}$ Jy.  The
normalization is set by imposing continuity and using $\eta_{1}=54.0$
Jy$^{\lambda_1-1}$ sr$^{-1}$.  The model relation is shown as the
solid line in figure~\ref{fig:n_s_norm}.  The counts were extended
below 10 $\mu$Jy based on the extrapolation suggested by
\markcite{win93}Windhorst et al. (1993).
For a typical SZ threshold $S_{d}\gg 10\mu$Jy, the results are not
sensitive to the parameters of this extrapolation.  Our total 4.86 GHz
intensity from radio sources is $i_{tot}(4.86\mbox{GHz})
\equiv
\int_{0}^{\infty} S ({dn}/{dS}) dS = 3.7 \times 10^{3}$ Jy sr$^{-1}$ or
equivalently 5.1 mK.

At frequencies above 4.86 GHz, we approximate the mean spectrum of the
sources by a power law, $S_{\nu} \propto \nu^{-\gamma}$. Based on 
the observed fluxes between 1.41--8.44 GHz,
\markcite{win93}Windhorst et al. (1993) estimate the median spectral
index of sources with $S_{4.86} \sim 0.1$ mJy to be $\gamma=0.35 \pm 0.15$.
Radio sources with $S_{4.86} \la 1$ mJy have a median angular size $\la
2^{\prime\prime}$ (cf. Fig. 2 in \markcite{win93}Windhorst et al.  1993).
In the FIRST radio survey, \markcite{whi97}White et al. (1997) found that
only $\sim20\%$ of all sources with $S(1.4\mbox{GHz}) \ga 1$ mJy have a
major axis $\ga 5^{\prime\prime}$.  The angular extent of sub--mJy radio
sources is thus much smaller than the Einstein angle of clusters with
observed arcs ($\alpha \sim 30^{\prime\prime}$; see
\markcite{lef94}Le F\`{e}vre et al. 1994), as well as the beam size
used in SZ measurements ($\ga 1'$; cf. \markcite{rep95} Rephaeli 1995).  
Thus, we can safely ignore the finite extent of the radio sources and 
treat them as if they were pointlike.

\section{The Lensing Effect}
\label{lensing_effect}
In a region of the sky where the magnification factor is $\mu$, the
apparent differential count of sources obtains the value $\left.
({dn}/{dS})\right|_{S} = \mu^{-2} \left.
({d\hat{n}}/{d\hat{S}})\right|_{S/\mu}$, where hat denotes unlensed
quantities. In particular, for a power law differential count
relation, $\left. ({d\hat{n}}/{d\hat{S}})\right|_{\hat{S}} \propto
\hat{S}^{-\lambda}$, the observed differential count is,
$\left. ({dn}/{dS})\right|_{S} \propto \mu^{\lambda-2} S^{-\lambda}$.  The
differential count therefore increases (decreases) as $\mu$ increases if
$\lambda$ is above (below) the critical slope $\lambda_{\rm crit} \equiv
2$. When $\lambda=\lambda_{\rm crit}$, lensing has no effect on the
apparent differential count. Interestingly, figure~\ref{fig:n_s_norm} shows
that the actual radio count slope oscillates around $\lambda_{\rm crit}$ for
fluxes $S_{4.86}\la 1$Jy.

In measurements of the SZ effect, discrete sources are typically
removed down to a given detection flux threshold, $S_{d}$. The mean
residual intensity $i(<S_{d})$ due to the superposition of all
undetected discrete sources with fluxes below $S_{d}$ is then assumed
to be equal to its sky--averaged value
\begin{equation}
\hat{i}(<S_{d})=\int_{0}^{S_{d}} d\hat{S}
\hat{S} \left. \frac{d\hat{n}}{d\hat{S}}
\right|_{\hat{S}} .
\end{equation}
However, the magnification due to lensing lowers the
unresolved intensity {\it systematically} relative to its sky--averaged
value and changes it to,
\begin{equation}
i(<S_{d}) = \int_{0}^{S_{d}} dS S \mu^{-2}
\left. \frac{d\hat{n}}{d\hat{S}} \right|_{S/\mu}
=\hat{i}(<S_{d}/\mu),
\end{equation}
where $\mu=\mu(\theta)$ is given by equation~(\ref{eq:mu_theta}).  Lensing
conserves the total intensity of the radio source background and merely
reduces the effective flux threshold for resolving sources by a factor
$\mu$.  The intensity offset due to lensing is then, $\Delta i_{\rm lens}
\equiv i(<S_{d})-\hat{i}(<S_{d})$. In the RJ regime, this can be expressed
more conveniently in terms of the brightness temperature difference, $\Delta
T_{\rm lens}=({c^{2}}/{2 \nu^2 k_{\rm B}}) \Delta i_{\rm lens}$, where
$k_{\rm B}$ is Boltzmann's constant.  For $\mu>1$ (i.e., $\theta >
\alpha/2$), the unresolved intensity is decreased, implying a negative
$\Delta T_{\rm lens}$, and so the SZ decrement in the RJ regime, $\Delta
T_{\rm SZ}$, is overestimated due to lensing. Note that for $\theta \gg
\alpha$, equation~(\ref{eq:mu_theta}) yields $\mu \approx 1+\alpha/\theta$
and $\Delta T_{\rm lens}\propto\theta^{-1}$.
\label{large_theta}

The effect of lensing on estimates of $H_{0}$ can be easily found from the
scaling, $H_{0} \propto (\Delta T_{\rm SZ})^{-2}$, where $\Delta T_{\rm
SZ}$ is the temperature offset produced by the SZ effect.  The small
systematic correction $(\Delta H_{0})_{\rm lens} = H_{0}(\mbox{true})-
H_{0}(\mbox{observed})$, which must be incorporated in order to compensate
for the lensing effect is, to leading order,
\begin{equation}
\frac{(\Delta H_{0})_{\rm lens}}{H_{0}} \approx 2 \frac{\Delta
T_{\rm lens}}{\Delta T_{\rm SZ}}.
\label{eq:dho}
\end{equation}
For $\theta \ga \alpha/2$ and the RJ spectral regime, both $\Delta T_{\rm
lens}$ and $\Delta T_{\rm SZ}$ are negative and so $(\Delta H_{0})_{\rm
lens}$ is positive.  The lensing correction will then tend to increase the
estimated value of the Hubble constant.

\section{Results}
\label{results}
\begin{figure}
\plotone{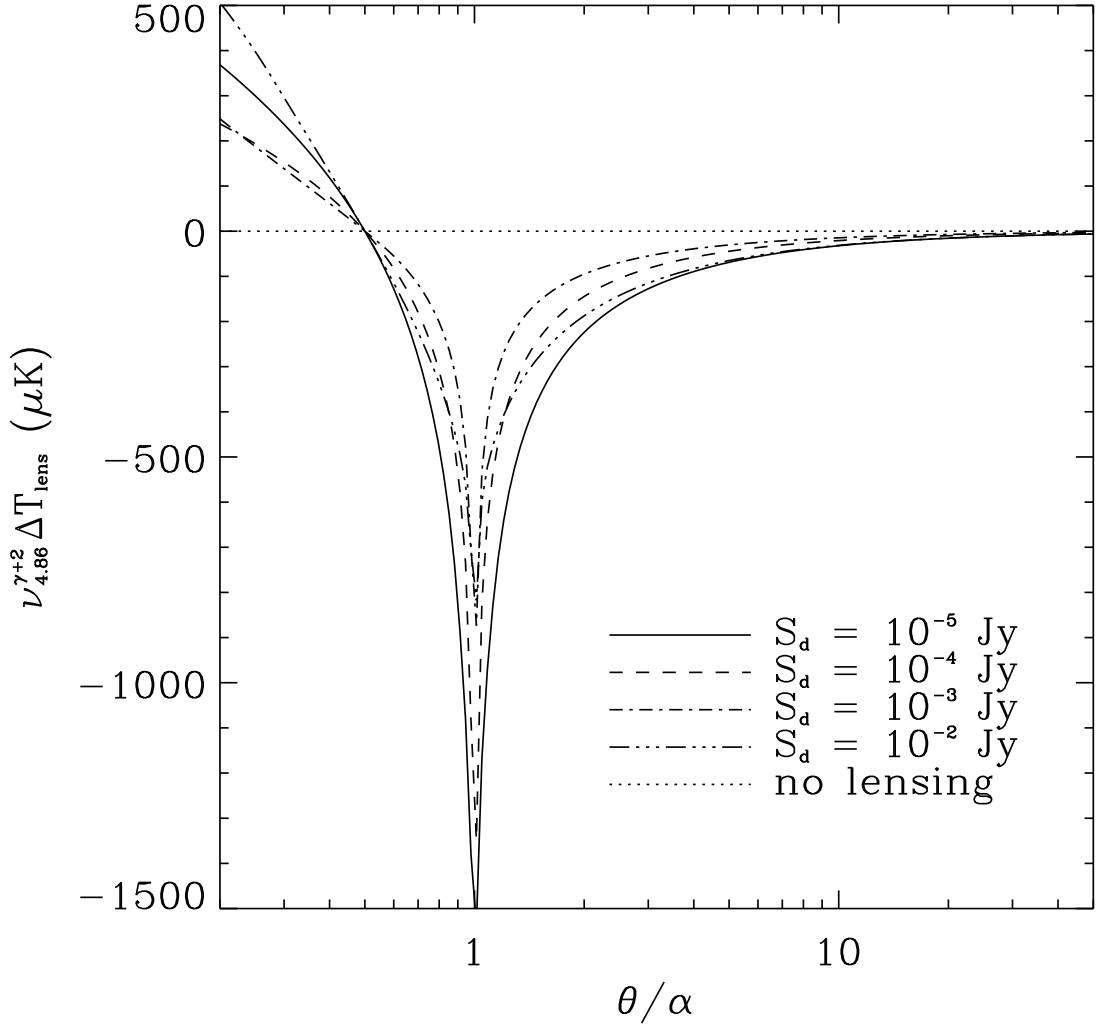}
\caption{Temperature offset $\Delta T_{\rm lens}$ induced by
gravitational lensing as a function of angular separation from the cluster
center. The offset is shown for several values of the source detection
threshold $S_{d}$. Values for $\Delta T_{\rm lens}$ and $S_{d}$ correspond
to a frequency $\nu=4.86$ GHz. The dependence of $\Delta T_{\rm lens}$ on
the Einstein angle of the cluster, $\alpha$, and on frequency, $\nu$, were
factored out. The parameter $\gamma$ is the mean spectral index of the
radio sources, and $\nu_{4.86}\equiv \nu/(4.86 \mbox{GHz})$.
\label{fig:dt_th_sd}}
\end{figure}

Figure~\ref{fig:dt_th_sd} shows $\Delta T_{\rm lens}$ as a function of
angular separation from the cluster center, $\theta$, for several values of
the detection threshold $S_{d}$. The values for $\Delta T_{\rm lens}$ and
$S_{d}$ correspond to a frequency of $4.86$ GHz. The dependence of $\Delta
T_{\rm lens}$ on the Einstein angle $\alpha$ and on frequency $\nu$ were
conveniently factored out.  The Einstein angles for clusters with observed
optical arcs are in the range of $10$--$50^{\prime\prime}$
(\markcite{lef94}Le F\`{e}vre et al. 1994).

The lensing decrement, $\Delta T_{\rm lens}$, shows a sharp peak near
the Einstein angle.  For $\theta \gg \alpha$, $\Delta T_{\rm lens}$ is
first weakened and then enhanced as $S_{d}$ varies from $10^{-2}$ to
$10^{-5}$ Jy. This is due to the fact that the count slope
$\lambda$ crosses the critical value $\lambda_{\rm crit}=2$ around $S_{d}
\sim 10^{-3}$ Jy (see Fig.~\ref{fig:n_s_norm}). The enhancement in
$\Delta T_{\rm lens}$ as $S_{d}$ decreases below $10^{-3}$ Jy occurs in
spite of the decrease in the unlensed intensity $\hat{i}(<S_{d})$ there.
At these fluxes, the removal of fainter radio sources paradoxically makes
the lensing decrement more pronounced. Note that because of the large shot
noise in the source counts (with an rms of $\sigma_{i}/i \approx 0.5$ in a
1 arcmin$^{2}$ cell for $S_{d}(4.86\mbox{GHz})=10^{-3}$ Jy), $\Delta T_{\rm
lens}$ will {\it not} necessarily be realized in each individual cluster.
The lensing induced decrement should be regarded as a systematic effect
that must be corrected for statistically, when a large sample of clusters
is considered. For observations with a large field of view,
the lensing signature might appear in the outer part of each individual
cluster.

As a specific example we consider A2218, an Abell richness class 4 cluster
at a redshift $z_{l}=0.175$, which shows several optical arcs
(\markcite{pel92}Pell\'{o} et al. 1992,
\markcite{leb92}Le Borgne et al. 1992).  Arc no. 359 in Pell\'{o} et
al. (1992) is separated by $20^{\prime\prime}.8$ from the central cD galaxy
and has a measured redshift of 0.702, close to the probable median redshift
of sub-mJy sources ($z_{s} \sim 0.5$--0.75; cf. \markcite{win93}Windhorst
et al.  1993).  We therefore model the cluster potential as a SIS with an
Einstein angle of $\alpha=20^{\prime\prime}.8$ for our radio sources (see
also
\markcite{mir95}Miralda-Escud\'{e} \& Babul 1995).
Interferometric imaging of the SZ effect in this cluster was performed by
\markcite{jon93}Jones et al. (1993) at 15 GHz, after the removal of point
sources with fluxes above $S_{d}(15\mbox{GHz})
\approx 1$ mJy. The restoring beam for their short baseline image had 
a FWHM of $129^{\prime\prime}\times 120^{\prime\prime}$.  The observed
angular dependence of $\Delta T_{\rm SZ}$ was fitted by 
a $\beta$--model,
\begin{equation}
\Delta T_{\rm SZ}(\theta)=\Delta T_{0}
  \left( 1 + \frac{\theta^2}{\theta_{c}^2} \right)^{1/2-3\beta /2}.
\label{eq:dt_sz}
\end{equation}
Acceptable $\chi^2$ values were obtained for different sets of parameters
ranging from $\beta\approx 0.6$, $\theta_{c} \approx 0^{\prime}.9$, and
$\Delta T_{0}
\approx 1.1$ mK, to $\beta \approx 1.5$, $\theta_{c} \approx 2^{\prime}.0$,
and $\Delta T_{0}=0.6$ mK.

\begin{figure}
\plotone{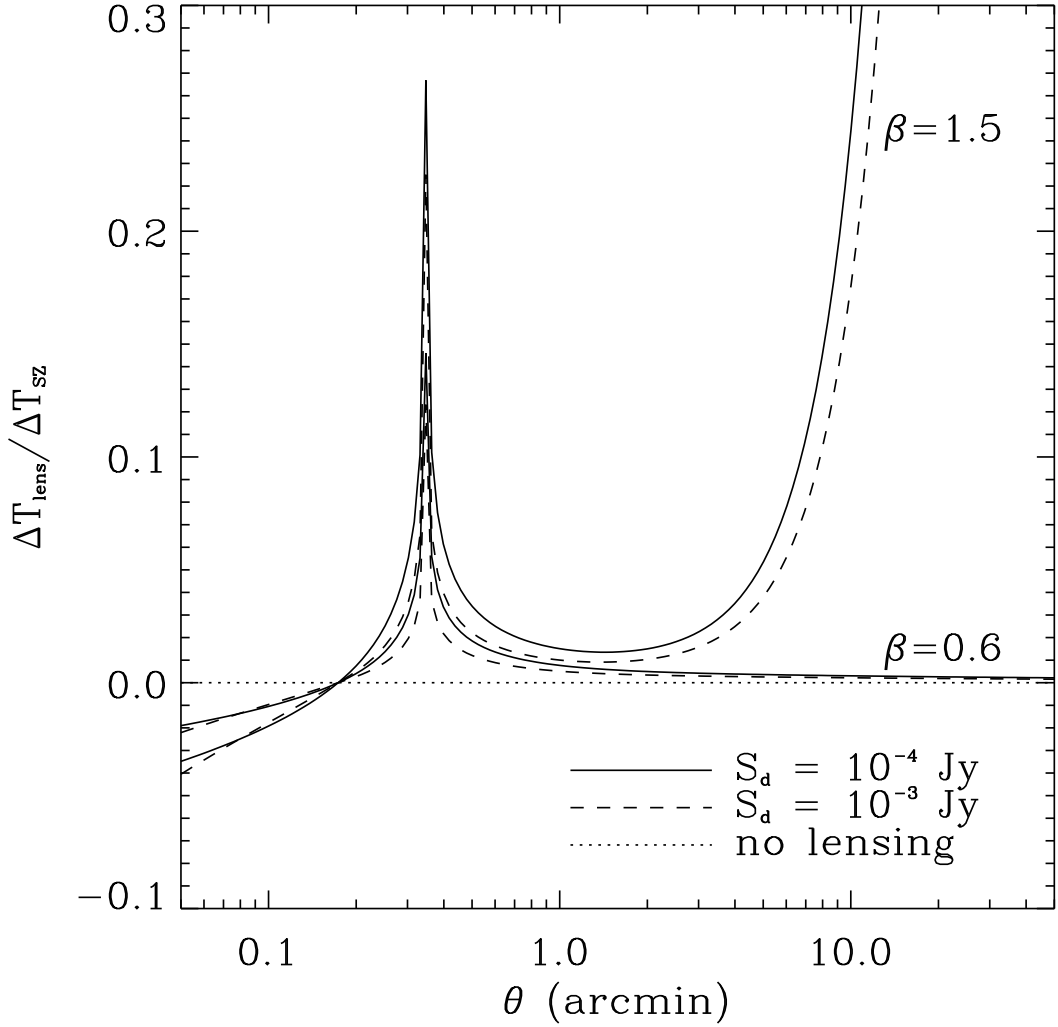}
\caption{Lensing effect in A2218.
The ratio of the temperature decrement induced by lensing
at 15 GHz, $\Delta
T_{\rm lens}$, to that induced by the SZ effect, $\Delta T_{\rm SZ}$, is
shown for two values of the flux detection threshold $S_{d}(4.86
\mbox{GHz})$. 
The decrement ratio is evaluated for the two extreme fits obtained by Jones
et al. (1993) to the observed radial dependence of $\Delta T_{\rm SZ}$.
The curves which converge (diverge) at large radii correspond to the
$\beta=0.6$ ($\beta=1.5$) fit.
\label{fig:dt_a2218}}
\end{figure}

Figure~\ref{fig:dt_a2218} shows the expected ratio $\Delta T_{\rm
lens}/\Delta T_{\rm SZ}$ for A2218 at 15 GHz, assuming a source
spectral index of $\gamma=0.35$. The ratio is shown for two values of
$S_{d}(4.86\mbox{GHz})$ and for the two extreme sets of fit parameters
for $\Delta T_{\rm SZ}(\theta)$. The sharp peak at
$\theta=20^{\prime\prime}.8$ reflects the enhancement in $\Delta
T_{\rm lens}$ around the Einstein radius (see
Fig.~\ref{fig:dt_th_sd}). For the $\beta=1.5$ model, $\Delta T_{\rm
lens}/\Delta T_{\rm SZ}$ diverges at $\theta \ga 2^{\prime}$.  If the
mass distribution follows the SIS profile, $\Delta T_{\rm lens}
\propto \theta^{-1}$ at $\theta \gg \alpha$. Since $\Delta T_{\rm SZ}
\propto \theta^{1-3\beta}$ for $\theta \gg \theta_{c}$
(cf. Eq.~[\ref{eq:dt_sz}]), the ratio $(\Delta T_{\rm lens}/\Delta T_{\rm
SZ}) \propto \theta^{3\beta-2}$ diverges at large radii if $\beta>2/3$. The
values of $\beta$ derived from X--ray observations of clusters have a large
scatter around a mean value of $\sim0.65$ (\markcite{sar88}Sarazin1988,
\markcite{jon84}Jones \& Forman 1984, Bahcall \& Lubin 1994). 
Weak lensing studies in the optical band could be used in conjunction with
X--ray observations to predict the relative radial behavior of the lensing
and SZ decrements in each individual cluster.

It is convenient to average the temperature offset over a circular ``top
hat'' beam of radius $\theta_{b}$ centered on the cluster center,
$\langle{\Delta T}(\theta_{b})\rangle \equiv 2 \theta_{b}^{-2}
\int_{0}^{\theta_{b}} \theta \Delta T(\theta) d\theta$. 
For the above model of A2218 with $S_{d}(4.86\mbox{GHz})=10^{-4}$ Jy, the
15 GHz mean temperature offsets due to lensing are $\langle{\Delta T_{\rm
lens}}\rangle \approx -28,-14,$ and $-1.6$ $\mu$K, for $\theta_{b}=0.4, 1$,
and $60$ arcmin, respectively.  The corresponding decrement ratios are
$\langle{\Delta T}_{\rm lens}\rangle/\langle{\Delta T}_{\rm SZ}\rangle
\approx 0.05, 0.03$, $0.18$, for the $\beta=1.5$ fit, and $0.03,0.02,0.002$
for the $\beta=0.6$ fit. The fractional correction to the Hubble constant
(Eq.~[\ref{eq:dho}]) is then $\Delta H_{0}/H_{0} \approx
6$--$10\%,4$--$6\%$, and $0.4$--$40\%$ for
$\theta_{b}=0.4, 1$, and $60$ arcmin, respectively, where the ranges
reflect the ambiguity in the fit parameters of $\Delta T_{\rm SZ}(\theta)$.

\section{Conclusions}
\label{conclusion}
We have shown that gravitational lensing of unresolved radio sources
leads to a systematic overestimate of the SZ temperature decrement at
angles $\theta>\alpha/2$. The amplitude of the lensing effect peaks
close to the Einstein angle of the cluster, $\alpha\sim 30^{\prime
\prime}$ (cf. Fig. 2).  While $\Delta T_{\rm SZ}$ is independent of
frequency in the RJ regime, the lensing decrement $\Delta T_{\rm
lens}\propto\nu^{-2-\gamma}$ (with $\gamma \approx 0.35$) is significant
only at frequencies $\nu \la 30$ GHz.  In clusters where the radial profile
of the gas pressure is steeper than that of the dark matter density (e.g.,
due to a gradient in the gas temperature), the ratio of the lensing to the
SZ decrement increases at large projected radii.  For observations of A2218
at 15 GHz with a source removal threshold of
$S_{d}(4.86\mbox{GHz})=10^{-4}$ Jy, $H_{0}$ could be overestimated by $\sim
0.4$--$40\%$, for a beam radius in the range of 0.4--60 arcminutes (cf.
Fig. 3).  The importance of the lensing effect will be enhanced in future
observations (including attempts to detect the {\it kinematic} SZ effect)
with greater sensitivity, higher angular resolution, and fainter source
removal threshold.

Lensing should also affect the power spectrum of microwave background
anisotropies on $\sim 1^\prime$ scales behind the cluster.  These
anisotropies are expected to originate primarily from the
Ostriker--Vishniac effect (Hu \& White 1996) and the cumulative SZ
effect of other background clusters (Colafrancesco et al.  1994,
\markcite{rep95}Rephaeli 1995).  Future SZ experiments might be
contaminated by noise from these fluctuations ($\Delta T/T \la 10^{-6}$),
especially in the outer parts of clusters.  However, since these diffuse
fluctuations will not be removed, lensing will conserve their net intensity
and will not systematically offset the SZ decrement as it does in the case
of discrete sources.

\acknowledgements We thank D. Helfand for useful comments on the
manuscript. This work was supported in part by the NASA grants NAG5-3085
(for AL) and NAGW2507 (for AR).

\end{document}